\documentclass[12pt]{article}
\usepackage{times}
\usepackage{geometry}
\usepackage{graphicx}
\usepackage[utf8]{inputenc}
\usepackage{amssymb} 
\usepackage{ragged2e}
\usepackage{pifont}

\begin{document}
\raggedright
\huge
Astro2020 Science White Paper \linebreak

The Dynamic Infrared Sky \linebreak
\normalsize

\noindent \textbf{Thematic Areas:} \hspace*{60pt} $\square$ Planetary Systems \hspace*{10pt} $\square$ Star and Planet Formation \hspace*{20pt}\linebreak
$\checkmark$ Formation and Evolution of Compact Objects \hspace*{31pt} $\square$ Cosmology and Fundamental Physics \linebreak
  $\checkmark$  Stars and Stellar Evolution \hspace*{1pt} $\square$ Resolved Stellar Populations and their Environments \hspace*{40pt} \linebreak
  $\square$    Galaxy Evolution   \hspace*{45pt} $\checkmark$             Multi-Messenger Astronomy and Astrophysics \hspace*{65pt} \linebreak
  
\textbf{Principal Author:}

Name: Mansi M. Kasliwal	
 \linebreak						
Institution:  Caltech
 \linebreak
Email: mansi@astro.caltech.edu
 \linebreak
Phone:  626-395-1575
 \linebreak
 
\textbf{Co-authors:} (names and institutions, alphabetical)
  \linebreak
Scott Adams (Caltech), 
Igor Andreoni (Caltech)
Michael Ashley (UNSW),
Nadia Blagorodnova (Radboud),
Kishalay De (Caltech),
Danielle Frostig (MIT),
Gabor Furesz (MIT),
Jacob Jencson (Caltech),
Matt Hankins (Caltech),
George Helou (IPAC),
Ryan Lau (JAXA),
Anna Moore (ANU),
Eran Ofek (Weizmann),
Rob Simcoe (MIT),
Jennifer Sokoloski (Columbia/LSSTCorp),
Jamie Soon (ANU),
Samaporn Tinyanont (Caltech),
Tony Travouillon (ANU)

\bigskip


\textbf{Abstract  (optional):}
Opening up the dynamic infrared sky for systematic time-domain exploration would yield many scientific advances. Multi-messenger pursuits such as localizing gravitational waves from neutron star mergers and quantifying the nucleosynthetic yields require the infrared. Another multi-messenger endeavor that needs infrared surveyors is the study of the much-awaited supernova in our own Milky Way. Understanding shocks in novae, true rates of supernovae and stellar mergers are some other examples of stellar evolution and high energy physics wherein the answers are buried in the infrared. We discuss some of the challenges in the infrared and pathfinders to overcome them. We conclude with recommendations on both infrared discovery engines and infrared follow-up machines that would enable this field to flourish in the next decade.   

\pagebreak

\setlength\parindent{1cm}

The dynamic infrared{\footnote{In this white paper, infrared refers to wavelengths between 1\,$\mu$m and 10\,$\mu$m}} sky is only now opening up for time-domain exploration. Many stellar fates shine the brightest in the infrared due to opacity or dust or temperature. Emission from neutron star mergers is longest lived and ubiquitous in the infrared as the bound-bound opacity of heavy elements pushes the peak of the emission to redder wavebands \cite{Barnes2013}. Emission from massive stars experiencing copious mass-loss could be self-obscured and better studied in the infrared \cite{Jencson19}. Emission from a Galactic supernova deep in the disk of the Milky Way may also be brightest in the infrared wavebands on account of line-of-sight extinction \cite{Adams13}. 

Despite the scientific value of infrared time-domain studies, the fundamental roadblocks have been the blindingly bright sky background from the ground (up to 250 times brighter than in optical wavelengths) and the narrow field-of-view of infrared cameras (smaller than a square degree). Infrared sensors are historically much more expensive than their optical CCD counterparts, and require lower operating temperatures, and thus sophisticated cooling methods. However, recent commercially available infrared sensor packages \cite{Simcoe19} combined with state-of-the-art software techniques, have enabled us to design a series of pathfinder surveyors in the infrared: Palomar Gattini-IR (25 sq deg field, 30\,cm aperture, J-band)~\cite{MooreKasliwal19} , WINTER (1.1 sq deg field, 100\,cm aperture, yJHs-bands)~\cite{Simcoe19} and DREAMS (3.75 sq deg field, 50\,cm aperture, yJHs-bands)~\cite{Soon18}. 


Below, we describe five science cases in more detail and conclude with some recommendations for the next decade.

\section{Neutron Star Mergers}

On August 17 2017, the groundbreaking discovery of both gravitational waves \cite{GW170817} and electromagnetic radiation \cite{MMA} from a binary neutron star (NS) merger marked the dawn of a new era in multi-messenger astrophysics. 
For the first time, we saw direct evidence of r-process nucleosynthesis, the process by which half the elements in the periodic table heavier than iron are synthesized. Heavy line blanketing from the large density of bound-bound transitions renders the opacity of r-process rich matter much higher than the conventional iron peak elements, shifting the emergent spectrum out of the optical bands and into the infrared. Therefore, the infrared data are the key to understanding the nucleosynthesis. Analysis of the infrared photometric evolution and the vivid broad features in the infrared spectroscopic sequence showed that at least 0.05 solar masses of heavy elements were synthesized by GW170817 \cite{Coulter17,Drout17,Evans17,Kasliwal17c,Smartt17,SoaresSantos2017,Cowperthwaite17,Arcavi2017}.
Combining the observed ejecta mass with rate estimates indeed gives numbers in the ballpark of the observed r-process abundances in the solar neighborhood. However, it is much debated whether the distribution of elemental abundances resembles the solar distribution. Evidence for the synthesis of the heaviest elements comes from late-time mid-infrared studies \cite{Kasliwal19}.  

Even as we celebrate 150 years since Mendeleev's periodic table, this first discovery opens up many questions for future discoveries to answer. Are NS- NS mergers the only sites of r-process nucleosynthesis? Do NS-NS mergers produce heavy elements in the same relative ratio as seen in the solar neighborhood? Are the heaviest elements in the third r-process peak, such as gold and platinum, synthesized? Which elements are synthesized in a NS-BH merger? 

To answer these questions by localizing additional gravitational wave events, there is an abundance of wide-field optical cameras being developed at all scales (including LSST). However, none of the NS-BH mergers and only a subset of NS-NS mergers may shine in the optical \cite{Kasen17}. Specifically, it is predicted that the optical emission is limited to polar viewing angles in NS-NS mergers where the velocities are relatively high or the opacities are relatively low due to neutrino irradiation. If the merger remnant collapses promptly to form a BH, the optical emission is suppressed. On the other hand, all NS-NS and NS-BH merger models predict that bright infrared emission from radioactive decay of heavy elements is ubiquitous and independent of mass ratio, remnant lifetime, viewing angle, opacity distribution and velocity distribution \cite{Kasen17}. Moreover, higher ejecta mass may imply that NS-BH mergers are especially luminous in the IR.

Thus, there is a critical need for a wide-field infrared transient survey that hunts for the elusive electromagnetic counterparts to gravitational waves. Both WINTER and DREAMS are designed to be sensitive enough to discover the electromagnetic counterpart to a neutron star merger as far out as the LIGO/Virgo gravitational wave horizon of 200\,Mpc.  As gravitational wave interferometers become more sensitive in the coming decade, even more powerful infrared discovery engines would be needed to localize neutron star mergers. 

\begin{figure*}[!hbt]
\centering
 \includegraphics[height=0.25\textwidth]{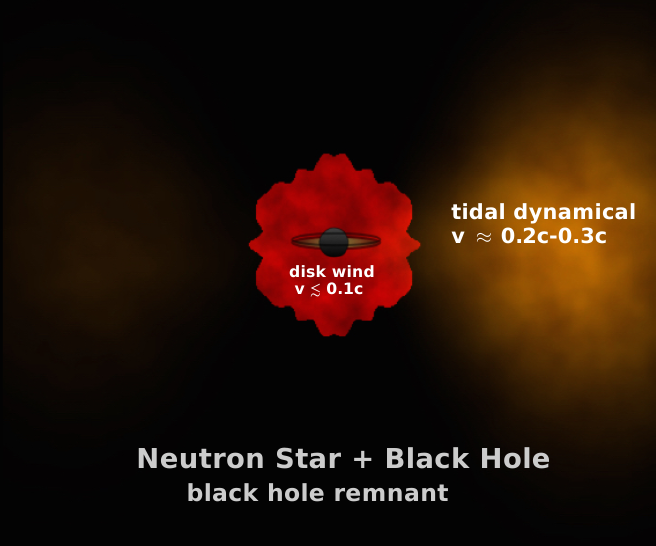} \includegraphics[height=0.27\textwidth]{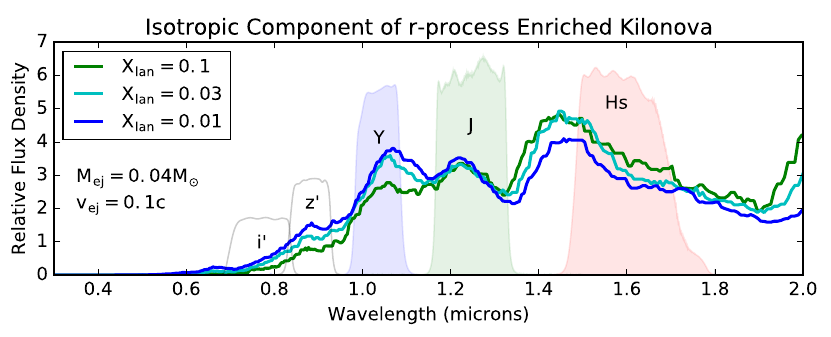} 
\caption{\footnotesize When a neutron star merges with another neutron star or a black hole, the emission peaks in the infrared wavelengths for a wide range of model parameters \cite{Kasen17}. The infrared filters for the proposed WINTER and DREAMS cameras are the yJHs-bands (shaded in color). Note contrast with the reddest filters (i,z) for optical CCD cameras.
\label{fig:winter}}
\end{figure*}


\begin{figure}[!hbt]
\includegraphics[height=0.3\textwidth]{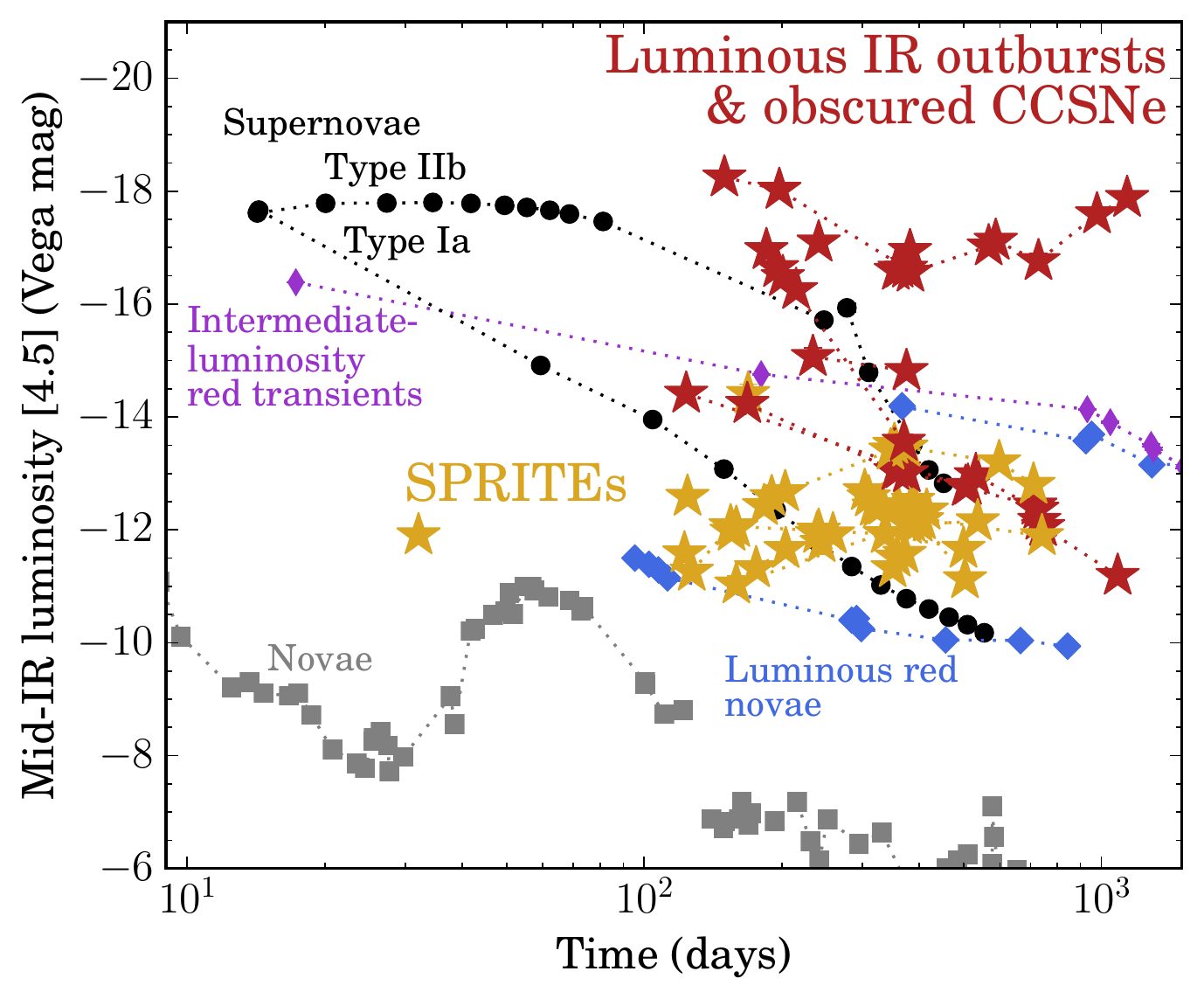} 
\centering
\caption
{\small \footnotesize {\it Left:} Phase space of infrared transients. Mid-IR light curves of various transients including SNe (black circles), novae (grey squares), stellar mergers (LRNe; blue diamonds) and  e-capture supernovae (ILRTs; purple diamonds). A new class of transients (SPRITEs, yellow stars; \cite{Kasliwal17a}), have luminosities intermediate between SNe and novae. The most luminous SPIRITS transients are likely obscured SNe (red stars; Jencson et al. 2017, 2018, 2019) 
\label{fig:sprites}
}
\end{figure}


\section{Obscured Supernovae}
A systematic transient survey with the Spitzer Space Telescope has revealed that even our inventory of supernovae in our backyard is woefully incomplete. The SPIRITS (SPitzer Infrared Intensive Transients Survey; \cite{Kasliwal17a}) survey imaged $\approx$200 nearby galaxies over a dozen times over 6 years and searched for transients. The visual extinction spans 1.5 to 9\,mag and the missed fraction of core-collapse supernovae within 35\,Mpc was 38.5\% \cite{Jencson17,Jencson18,Jencson19}. Increasing the sample size of such infrared-only supernovae could help reconcile claimed discrepancies between the cosmic star formation and the optically constrained core-collapse supernova rates. Some supernovae show hints of being obscured by the molecular cloud that formed the star \cite{Jencson18}. Moreover, some transients underwent multiple, infrared outbursts indicative of violent mass-loss histories leading up to explosion (with no optical detections during any of these outbursts) \cite{Jencson19}. If future infrared surveys show this is common, it could be symptomatic of a separate class of progenitor systems that are subject to extreme environmental conditions at the time of explosion. 

\section{Exotic Transients}
SPIRITS has also uncovered 64 unusual infrared transients in the luminosity gap between novae and supernovae, dubbed ``SPRITEs" (eSPecially Red Intermediate-luminosity Transient Events) \cite{Kasliwal17a}. SPRITEs occur in grand spirals, emit predominantly in the infrared and have no optical counterparts whatsoever. These transients are neither detected in our many ground-based concomitant optical surveys nor are they detected in deep HST imaging. Infrared colors suggest very cold effective blackbody temperatures spanning 350 to 1000 K. These transients cannot be classical novae as the mid-infrared luminosities are much higher than Eddington. The colors and luminosity cannot be explained simply as extremely dust obscured SNe. The photometric evolution is diverse and spans slow rise, flat evolution and fast decline. We speculate these may represent diverse physical origins such as the birth of a massive star binary, electron- capture induced collapse of an extreme AGB star, outflows in mass-losing binaries, and stellar mergers.

\section{Classical Novae}
Classical novae have been studied over the past century and thought to be thermonuclear runaways on the surface of the white dwarf that power a transient flash that is a million times the luminosity of the sun. However, just five years ago, the discovery by the Fermi satellite that many classical novae produce $\gamma$-ray emission has revealed that shocks are essential to novae \cite{Ackermann14}. The shocks likely arise as a wind from the eruptive white dwarf collides with an equatorial torus of material. It appears that powerful shocks plowing into a likely dusty torus influence every aspect of the nova, from emission at radio through $\gamma$-ray wavelengths to the morphology of the nova remnant \cite{Derdzinski17}. These shocks might even lead to high density regions that could form dust despite the intense flux of ionizing radiation from the white dwarf.
Thus, combining the infrared light curves of novae discovered by Palomar Gattini-IR with $\gamma$-ray and radio data would address this shock-dust conundrum. 
[For more details, see white paper by Chomiuk et al.]

\section{The Next Supernova in the Milky Way}
Despite our many multi-messenger facility advances in the past century, we have not yet witnessed a supernova in our Milky Way. In the upcoming decade, if a Galactic supernova is detected by neutrino detectors and perhaps even, gravitational wave detectors, it would be a tragedy to miss the electromagnetic counterpart on account of dust extinction. Extinction is many magnitudes lower in the infrared than the optical. The wide-field capability of the Palomar Gattini-IR system would be sensitive to nearly the entire distribution of progenitors, shock breakouts and supernovae from a star located anywhere in our galaxy (compare to distributions in \cite{Adams13}). Building a global network of such systems spread over various latitudes and longitudes will help mitigate obstacles such as the sun, elevation limit and weather.  

\section{Recommendations for 2020s}
In summary, there is tremendous untapped scientific potential in exploring the dynamic infrared sky. The next discovery frontier is best attained by building a sensitive wide-field imager ($\sim$10 square degree field-of-view) at a location unhindered by sky background i.e. the North Pole or South Pole or Space. Time-domain discovery in the infrared requires balancing survey area, survey depth and survey cadence. WFIRST is exquisitely sensitive but much too narrow (quarter square degree), much too slow a cadence (every five days) and limited to 2\,$\mu$m as currently designed. NEOCAM is exquisitely wide area but somewhat shallow and moderate cadence [see WP by Ciardi et al.]. A dedicated time-domain surveyor on a polar location could optimize all three time-domain axes. Specifically, it has been shown that the sky background at 2.35\,$\mu$m is remarkably (50 times) lower from the cold polar regions, such as the Antarctic high plateau bases at the South Pole, Dome A and Dome C, than any other location on earth. A wide-field infrared telescope (such as the Turbo Gattini-IR concept) could easily survey $\approx$10,000 sq. deg. to a staggering depth of 20 mag AB every 2 hours at 97\% efficiency. [For comparison, pathfinder Palomar Gattini-IR covers $\approx$10,000 sq. deg. to 16\,mag AB every night and WINTER/DREAMS will cover $\approx$10,000 sq. deg. to 18\,mag AB every three nights].

Discovery of infrared transients is only the first step. Detailed spectroscopic characterization is essential to get to the core of the astrophysics. With more prolific discovery engines, the infrared spectroscopic classification capability on ground-based 4m-class to 8m-class telescopes would need to be enhanced to efficiently triage the events, build statistically robust samples and identify the rarest transients for detailed characterization .  The proposed space-based Time-domain Spectroscopic Observatory (TSO) dedicated to rapid transient classification could also efficiently and sensitively classify transients. 

Beyond classification is detailed characterization.  Late-time Spitzer mid-infrared detections of GW170817 suggest that a handful of isoptopes with half-life around 14\,days could dominate the emission at this phase (e.g. $^{140}$Ba, $^{143}$Pr, $^{147}$Nd, $^{156}$Eu, $^{191}$Os, $^{223}$Ra, $^{225}$Ra, $^{233}$Pa, $^{234}$Th) \cite{Kasliwal19}. Infrared spectroscopy with the James Webb Space Telescope would be essential to characterize the full spectral energy distribution (e.g., MIRI and NIRSPEC spectra out to 12$\mu$m), identify elements and disentangle multi-component contributions. The next generation of extremely large telescopes (TMT, GMT and e-ELT) would also be invaluable especially to obtain spectra in the late, nebular phase to directly unravel the chemical composition.

\pagebreak

\bibliography{ref2}
\bibliographystyle{Science}

\end{document}